\documentclass{article}
\usepackage{graphicx} 
\usepackage{amsmath}
\usepackage{relsize}   
\usepackage[utf8]{inputenc}
\usepackage{float}
\usepackage{amsmath}
\usepackage{setspace}
\usepackage{upgreek}
\usepackage{siunitx} 
\usepackage{graphicx}
\usepackage{amsmath, amssymb}
\usepackage{authblk}
\usepackage{url}
\usepackage{caption}

\usepackage[margin=1in]{geometry} 

\title{Equivalence of Optical Theorems}

\author[1] {Edwin A. Marengo\thanks{Corresponding author: e.marengo@northeastern.edu}}
\author[1] {Mohammadrasoul Taghavi}
\affil[1]  { Department of Electrical and Computer Engineering, Northeastern University, Boston, Massachusetts 02115, USA}

\date{}
\begin{document}

\maketitle
\begin{abstract}
    We demonstrate, in the full vector formulation of electromagnetic fields, that the well-known optical theorem pertinent for the characterization of a scatterer's extinction power and associated cross section can be expressed in a multitude of alternative equivalent forms.  These alternatives involve different forms of projective field measurements or detectors. The inherent nonuniqueness of such optical-theorem-based detectors stems from the nonuniqueness of an associated inverse source problem, and can be interpreted via well-known equivalence principles. Some of the multiple ways in which the extinction of power due to the interaction of a scattering body with a probing field can be measured remotely are derived and interpreted for a number of canonical frameworks. This includes detectors and their corresponding optical theorems synthesized in the contexts of surface-confined sensors for near-field sensing, surface sensors based on backpropagation-based imaging, a number of planar aperture realizations dealt with through classical diffraction theory, as well as detectors based on multipole representations. General aspects of the derived optical theorems are discussed in the context of envisioned practical applications. 
\end{abstract}

\section{Introduction}

The optical theorem is a fundamental result in wave physics, which enables the measurement of the total power removed from an incident wave by a medium perturbation or scatterer (customarily known as  ``the extinction power'') by means of a suitable projective  measurement of the corresponding scattered field. In the classical version of this result, pertinent to plane-wave illumination in an unbounded homogeneous medium,  the relevant datum carrying all the information about the power budget of the scattering phenomenon is the forward scattering amplitude 
 \cite{born2013principles, newton2013scattering}. 
This interesting link between the extinction power and the forward scattering amplitude, which is essentially a measurable quantity, constitutes a powerful tool for wave-matter interaction analysis. 

In the past two decades, researchers have significantly expanded the theorem's foundational scope. One notable line of progress is the generalization of the optical theorem for the case of non-plane wave incident waves  \cite{gouesbet2023failures,gouesbet2009optical, lock1995failure, krasavin2018generalization,carney1997statistical,rondon2017generalized, mitri2015generalization, martin2024acoustic,rondon2022analytical,eremin20177generalized}. In another work, the optical theorem has been extended to non-diffracting beams and applied to the calculation of radiation force and torque \cite{zhang2013optical}. Another successful line of work has been the generalization of this theorem to nonhomogeneous media \cite{dacol2005generalized, lytle2005generalized, wapenaar2012unified}. Also, various adaptations of the optical theorem have been explored in the context of wave scattering within transmission lines \cite{marengo2014optical}, transient fields \cite{karlsson1998time,karlsson2002some, vstumpf2014time, marengo2016generalized}, anisotropic and nonlinear media \cite{marengo2012new, li2015optical}, cloaking \cite{slovick2022inverse}, and magnetodielectric media \cite{alu2010does}. Yurkin {\em et al.} \cite{moskalensky2019energy} provided a rigorous volume-integral-equation-based-formulation of the optical theorem for source-induced fields, clarifying power flows in complex scattering environments. A generalized optical theorem was developed for cylinders in absorbing media, where host absorption was shown to induce negative extinction and suppress resonance features \cite{zhang2022optical}. Small {\em et al.} \cite{small2013generalization} extended the optical theorem formulation to particles embedded at planar interfaces by establishing boundary conditions critical for extinction measurements. In another interesting work, Zhu {\em et al.} \cite{zhu2023multipole,zhu2023quantitative} employed the optical theorem to compute the extinction cross section of chiral metamolecules directly from the induced polarization rather than from far-field scattering, in order to analyze the contribution of electric, magnetic, and higher-order multipoles to the observed circular dichroism. Also, numerous works have explored generalizations of the optical theorem to account for multipole excitations \cite{eremin2021generalization,eremin2017generalization, eremin2017generalization2, eremin2016generalization}. 
In another work, Li {\em et al.} \cite{li2025understanding} have shown that the total scattering cross section of ellipsoids approximates their surface area in the long-wavelength limit, extending the optical theorem to more complex shapes. 
Gulyaev {\em et al.} \cite{gulyaev2005optical} extended the optical theorem to include evanescent waves where it has been shown that scattering by dielectric structures, analogous to quantum tunneling, can generate a propagating energy flux in the direction of decay. Recently, the optical theorem has been employed in digital holography to estimate the extinction cross section of both spherical and non-spherical particles from interference fringe patterns \cite{berg2017measuring, berg2023measuring, berg2024measuring}. More recent developments have extended this concept to complex scattering environments, enabling real-time, single-pixel detection and tracking of particles, and highlighting the need for generalizable optical theorem formulations applicable to arbitrary fields and geometries \cite{taghavi2025optical}.

In this paper, we revisit the optical theorem in the full-vector, electromagnetic context, and present a series of alternative, equivalent formulations. We build on prior work on the corresponding scalar-based formulation \cite{marengo2015nonuniqueness} where the intrinsic nonuniqueness of the sensing modes (called ``optical theorem detectors'' or ``OT detectors'') required for sensing of the extinction power is discussed. As in that paper, in the present full-vector extension, we show that such OT detectors are intrinsically nonunique and take instead a multiplicity of equivalent forms. Alternative forms of such OT detectors are derived in the paper, including exact forms that apply in the near field region of the probing source as well as approximate versions that apply under mild conditions but lack validity in the presence of near field evanescent components. Both are relevant in practice within their respective domains of validity. We demonstrate OT detector realizations pertinent to different sensing configurations, 
ranging from planar to cylindrical to spherical geometries. The demonstrated alternative formulations of the optical theorem facilitate practical implementations in different sensing setups, including complex geometries, and arbitrary probing sources, and are also expected to be particularly valuable in scenarios where classical far-field assumptions break down, such as near-field sensing. The framework introduced here supports practical implementations in power budget estimation, noninvasive imaging, inverse scattering, and modal characterization of structured media. Additionally, the multipolar and surface-based realizations offer new tools for analyzing wave interactions in photonic systems, antenna arrays, and scattering environments that can extend the reach of optical-theorem–based methods in both theoretical and applied domains. It is also shown that the results corresponding to these alternative frameworks exhibit certain common features despite their differences in terms of the adopted functional representation. Corollaries emerge from the derived results which are briefly discussed and have potential implications in cross-field sensing and sensing with intensity-only data, as well as in the characterization of the scattering response of certain classes of scatterers, e.g., lossless, passive, etc., and in the application of the optical theorem to scatterer detection and estimation problems. 

The rest of the paper is organized as follows. Section 2 presents an overview of the required electromagnetic scattering principles. Section 3 provides the main theoretical results, involving alternative equivalent realizations of OT detectors and their respective expressions for the extinction power. We also highlight commonalities of these alternative forms alongside interesting implications and practical applications of the derived results. Section 4 provides concluding remarks. 

\section{Background}

We begin the theoretical developments with an overview of the relevant energy conservation principles, upon which the following optical theorem results are based. We consider a homogeneous, lossless medium having permittivity and permeability $\epsilon$ and $\mu$, respectively. In this framework, electromagnetic sources and fields having suppressed time-dependence $e^{-i\omega t}$
are related by the relevant Maxwell's equations: 
\begin{eqnarray}
    \nabla\times {\bf E}({\bf r}) &=& i\omega\mu {\bf H}({\bf r}) - {\bf M}({\bf r}) \nonumber \\ 
    \nabla\times {\bf H}({\bf r}) &=& -i\omega\epsilon {\bf E}({\bf r}) + {\bf J}({\bf r}) 
    \label{eq_may24_2025_1}
\end{eqnarray}
where ${\bf r}$ denotes position, ${\bf E}$ is the electric field, ${\bf H}$ is the magnetic field, ${\bf J}$ is the electric current density, and ${\bf M}$ is the magnetic current density. If the sources ${\bf J}, {\bf M}$ are confined within a given region of compact support, say a region (e.g., a volume) $V$ that is bounded by $\partial V$, then the average power $P_d$ delivered by the sources to the electromagnetic field is given by Poynting's theorem, in particular
\begin{equation}
    P_d=-\frac{1}{2}\Re \int_V d{\bf r} \left ({\bf J}^*\cdot {\bf E} + {\bf M} \cdot {\bf H}^* \right ) = \frac{1}{2}\int_{\partial V} dS  \hat{\bf n}\cdot\left ( {\bf E}\times {\bf H}^*\right )  = P_o \label{eq_may24_2025_2}
\end{equation}
where $\Re$ denotes the real part, $dS$ denotes the differential element in the boundary $\partial V$, $\hat{\bf n}$ denotes the unit vector in the direction of the outward normal to the surface at the differential element position, and 
where in view of the lossless nature of the medium $P_d$ is equal to the average power $P_o$ exiting the boundary $\partial V$. 
The volume integral in (\ref{eq_may24_2025_2}) is representative of the (energy) interaction of the source and the field it generates. More generally, the power $P_{1\rightarrow 2}$  put by a source ``1'' (${\bf J}_1,{\bf M}_1$) of support $V$ on an electromagnetic field ``2'' (${\bf E}_2,{\bf H}_2$) is given by 
\begin{equation}
    P_{1\rightarrow 2}=-\frac{1}{2}\Re \int_V d{\bf r} \left ({\bf J}_1^*\cdot {\bf E}_2 + {\bf M}_1 \cdot {\bf H}_2^* \right ).
\label{eq_may24_2025_3}
\end{equation}

Scattering involves the probing of a medium perturbation or scatterer using incident waves, say electric (${\bf E}_i$) and magnetic (${\bf H}_i$) fields that are excited or generated in the relevant background medium, in the absence of the scatterer. A question of fundamental importance is the determination of the amount of extinction power, i.e., the quantification of the power taken away by the scatterer, from the probing or incident field. Equation (\ref{eq_may24_2025_3}) provides a framework to measure the extinction power, for a scatterer in $V$, wherein the pertinent field from which radiation power is taken is the incident field. Then we substitute ${\bf E}_2 \rightarrow {\bf E}_i$ and ${\bf H}_2 \rightarrow {\bf H}_i$. In addition, the relevant sources ${\bf J}_1, {\bf M}_1$ acting as power sinks are the sources ${\bf J}_s , {\bf M}_s$ that are induced in the scatterer in response to said excitation. Adopting these substitutions in (\ref{eq_may24_2025_3}) we find that the (extinction) power $P_e$ taken away from the probing or incident waves by the appearance (within the given background medium) of the scatterer in question is given by 
\begin{equation}
    P_e = \frac{1}{2}\Re \int_{V} d{\bf r } \left ({\bf J}_s^*\cdot {\bf E}_i + 
    {\bf M}_s \cdot {\bf H}_i^* \right ).
\label{eq_may24_2025_3b}
\end{equation}

The pertinent question is whether the extinction power $P_e$ in (\ref{eq_may24_2025_3b}) can be determined from measurements of the scattered field ${\bf E}_s,{\bf H}_s$ carried out outside the region $V$ where the scatterer is located, i.e., for ${\bf r}\in \bar V=\mathbb{R}^3\setminus V$. The most general such measurement takes the form of a projection of the scattered electromagnetic field ($ {\bf E}_s,{\bf H}_s$) onto modes ``$m$'' defined by the functions (${\bf I}_m,{\bf K}_m$): 
\begin{equation}
    w= \int_{\bar V} d{\bf r} \left ( {\bf I}_m\cdot {\bf E}_s - {\bf K}_m \cdot {\bf H}_s \right ) .\label{eq_may24_2025_6}
\end{equation}
Moreover, in view of (convolution-type) reciprocity, it follows that the general measurement in (\ref{eq_may24_2025_6}) can correspond to the interaction integral in (\ref{eq_may24_2025_3b}), 
i.e., 
\begin{equation}
    w = \int_V d{\bf r} \left ({\bf J}_s^*\cdot {\bf E}_i + 
    {\bf M}_s \cdot {\bf H}_i^* \right ) ,\label{eq_may24_2025_10}
\end{equation}
as long as the sensing modes or detectors ${\bf I}_m,{\bf K}_m$ are such that they generate, when they act as sources, the complex conjugated (c.c.) form of the incident fields ${\bf E}_i^*, -{\bf H}_i^*$ in the region of support $V$ or, more precisely, in the region (contained within $V$) where the scatterer resides. Henceforth we refer to sensing modes or detectors exhibiting this specific requirement as OT detectors, and we denote them sometimes as ${\bf J}^{(OT)}$ and ${\bf M}^{(OT)}$ in place of ${\bf I}_m$ and ${\bf K}_m$, respectively. In this notation, we obtain from eqs.(\ref{eq_may24_2025_3b},\ref{eq_may24_2025_6},\ref{eq_may24_2025_10}) the following general expression from which all the equivalent forms of optical theorem derived in this work follow:
\begin{equation}
P_e = \frac{1}{2}\Re (w) 
    \label{eq_may24_2025_15}
\end{equation}
where \begin{equation}
    w = \int_{\bar V} d{\bf r} \left [{\bf J}^{(OT)}\cdot {\bf E}_s -
    {\bf M}^{(OT)} \cdot {\bf H}_s \right ] \label{eq_may24_2025_10b}
\end{equation}
where the OT detectors ${\bf J}^{(OT)}, {\bf M}^{(OT)}$ generate, in radiation, the c.c. version of the probing fields interrogating the scatterer. We highlight in this work the fact that there is an inherent nonuniqueness of such detectors, which stems from the nonuniqueness of a companion inverse problem. This renders a certain flexibility in the practical realization of these sensors, enabling different sensing configurations as applicable to different sensing scenarios and constraints. We demonstrate in the following a number of viable realizations of such detectors that can be applied in different sensing topologies and geometries, and the corresponding results shed further insight on the information content linked to such measurements.

\section{Main Results and Discussion}
\subsection{Surface Optical Theorem Detectors}
\label{subsection_surface_source}

In the following, we denote $\tau_0$ as the scatterer's support, and $\tau\supseteq\tau_0$ as the region of interest (ROI) for which the derived OT detectors apply. The latter region is a volume bounded by the surface $\partial \tau$. Let $V$ be a volume enclosing $\tau$ ($\tau\subseteq V$) that is bounded by the surface $\partial V$, as depicted in Fig.~1. 
We discuss next optical theorems in which the source of the probing field (${\bf E}_i,{\bf H}_i$) is outside $V$ (meaning here exterior of $V$ and boundary $\partial V$), whereby the optical theorem sensor can be synthesized with surface sources confined within $\partial V$. In this physical situation, probing fields are applied to region $V$ and sensors are placed at the boundary $\partial V$ of the region for the purpose of measuring the scattering cross section of scatterers that are contained in the ROI $\tau\subseteq V$. As indicated earlier, the required optical theorem sensors must be such that they generate, in emission, the time-reversal or c.c. version of the probing fields in the ROI. This means that if the ROI is probed with incident fields ${\bf E}_i,{\bf H}_i$, then the OT detector must generate, in emission, electric and magnetic fields ${\bf E}_i^*$ and $-{\bf H}_i^*$, respectively.

A simple way to synthesize the required sensors is via the surface equivalence principle, and associated Love's equivalence principle (\cite{balanis2024balanis}, page 329), whereby arbitrary fields ${\bf E}, {\bf H}$ are generated in $V$ with surface sources defined by 
\begin{eqnarray}
    {\bf J}_s & = &-\left(\hat{\bf n}\times {\bf H}\right ) \delta_s({\bf r}) \nonumber  \\
    {\bf M}_s &=& \left (\hat{\bf n}\times {\bf E}\right ) \delta_s ({\bf r}) \label{eq_april22_2025_1}
\end{eqnarray}
where $\hat{\bf n}$ is the unit vector in the direction of the outward normal to surface $\partial V$, ${\bf J}_s$ and ${\bf M}_s$ are, respectively,  an electric surface source and a magnetic surface source, supported in the surface $\partial V$, and $\delta_s$ denotes the Dirac single layer singularity or singlet in that bounding surface. Therefore, it follows readily that surface sources defined by 
\begin{eqnarray}
    {\bf J}_s^{(OT)} & = &(\hat{\bf n}\times {\bf H}_i^* )\delta_s({\bf r}) \nonumber  \\
    {\bf M}_s^{(OT)} &=& (\hat{\bf n}\times {\bf E}_i^* )\delta_s ({\bf r}) \label{eq_april22_2025_1b}
\end{eqnarray}
generate, within $V$, the c.c. fields ${\bf E}_i^*, -{\bf H}_i^*$, and are therefore the sought-after OT detectors based on this principle. In particular, it follows from eqs.(\ref{eq_may24_2025_15},\ref{eq_may24_2025_10b},\ref{eq_april22_2025_1b})
that
for any scatterer of support $\tau\subseteq V$, the following projective measurement carries all the power budget information, giving for instance the sought-after extinction power $P_e$ associated with the scatterer upon excitation by the given probing field: 
\begin{equation}
    P_e = \frac{1}{2}\Re(w)\label{eq_april25_2025_2}\end{equation}
    where
\begin{equation}
    w = \int_{\partial V} dS  \left [ {\bf E}_s \cdot (\hat{\bf n}\times {\bf H}_i^*) - (\hat{\bf n}\times {\bf E}_i^*)\cdot {\bf H}_s \right ]\label{eq_april22_2025_2}
    \end{equation}
where $dS$ denotes surface differential element and ${\bf E}_s,{\bf H}_s$ are the corresponding scattered fields. It follows from a well-known vector identity that the result eq.(\ref{eq_april22_2025_2}) can be expressed also as \begin{equation}
    w  = - \int_{\partial V} dS \left [ \hat{\bf n} \cdot \left ( {\bf E}_s \times {\bf H}_i^* + {\bf E}_i^* \times {\bf  H}_s \right ) \right ]. \label{eq_april22_2025_3}
\end{equation}

\begin{figure}
    \centering
    \includegraphics[width=0.8\linewidth]{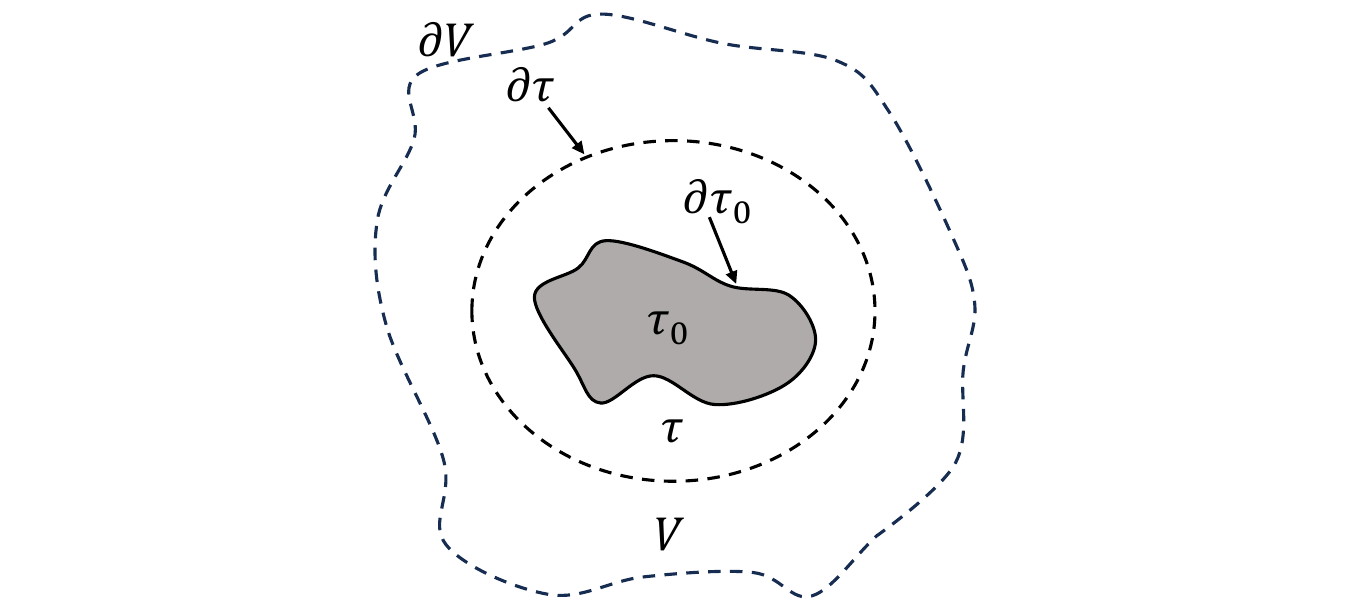}
    \caption{Illustration of the relations between the scatterer support $\tau_0$, the ROI $\tau\supseteq \tau_0$, and the sensing domain $\partial V$ adopted in the discussion of the OT detectors based on surface source realizations.}
    \label{fig}
\end{figure}

Importantly, the form of OT detector defined in eqs.(\ref{eq_april22_2025_1b},\ref{eq_april25_2025_2},\ref{eq_april22_2025_2},\ref{eq_april22_2025_3}) is valid for an arbitrary volume $V$, as long as the probing fields are generated by sources located outside $V$,  as illustrated in Fig.~2, and has applicability for the sensing of the extinction in the near field. 
%We discuss later specialized versions of this general result that apply to the far zone, as %well as for specialized geometries such as planar surfaces. 
\begin{figure}
    \centering
    \includegraphics[width=0.8\linewidth]{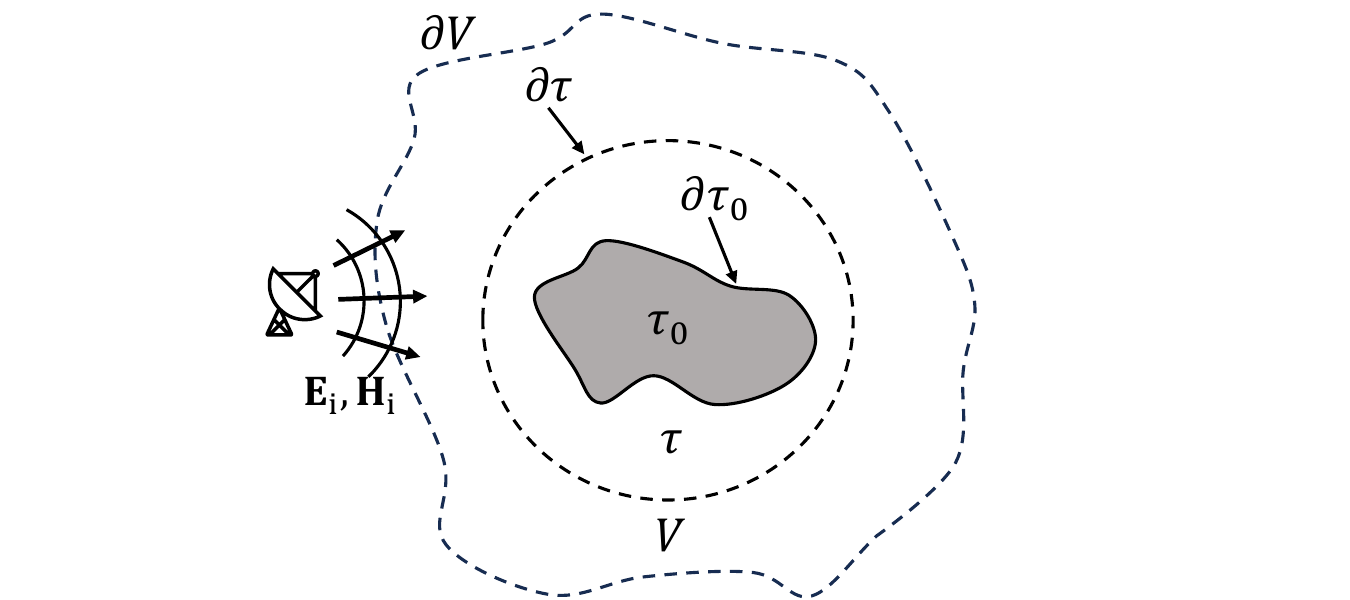}
    \caption{General topology for application of the OT detector based on surface sources derived from the surface equivalence principle. The probing fields can be generated by arbitrary sources outside $V\supseteq \tau$, and this includes near fields carrying evanescent content there.}
    \label{fig:enter-label}
\end{figure}
\begin{figure}
    \centering
    \includegraphics[width=0.8\linewidth]{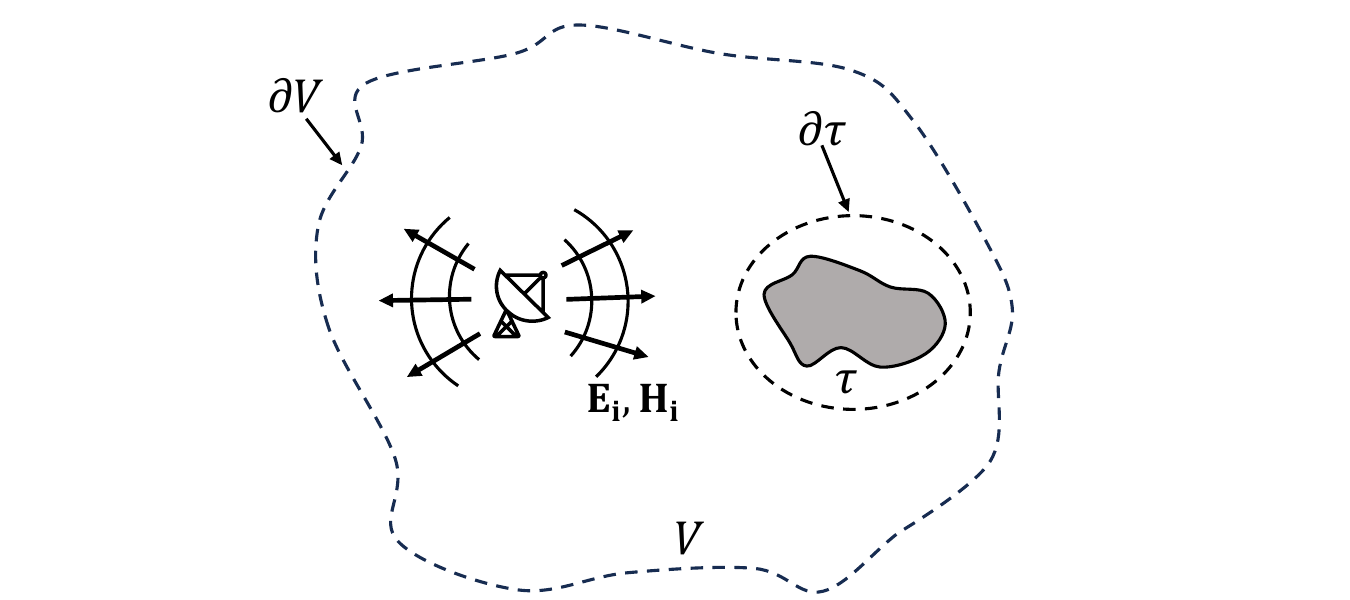}
    \caption{Topology for the backpropagation-based surface source form of the OT detector, in which the source of the probing fields resides in the same domain $V\supseteq \tau$, and at a sufficiently far distance (a few wavelengths) away from the ROI $\tau$ so as to lack significant evanescent content in $\tau$.}
    \label{fig:enter-label}
\end{figure}
\subsection{Backpropagation-based Surface Source Realization}
\label{subsection_backpropagation}

In the prior scenario the incident electromagnetic field is produced by sources located outside $V$ (see Fig.~2). Another important scenario is the case in which the source of the incident field is in the interior of the volume $V$ bounded by the sensing surface $\partial V$, as illustrated in Fig.~3. This case gives rise to further OT detector variants. In this case the  fields ${\bf E}_i , {\bf H}_i$, in $V$, and the corresponding c.c. forms (${\bf E}_i^*, -{\bf H}_i^*$), do not obey the homogeneous form of Maxwell's equations. Consequently, we cannot in general reproduce the c.c. fields in $V$ via sources supported outside $V$. However, we can approximate the c.c. field in the ROI $\tau\subseteq V$ if the probing source is far from $\tau$, so that the fields ${\bf E}_i, {\bf H}_i$ in the scattering region have limited evanescent components. In that case we can generate fields that are a good approximation (in $\tau$) of the c.c. fields via the following surface source in $\partial V$:
  \begin{eqnarray}
    {\bf J}_s^{(OT)} &=& (\hat{\bf n}\times {\bf H}^{*}_i)\delta_s({\bf r}) \nonumber \\
    {\bf M}_s^{(OT)} &=& (\hat{\bf n}\times {\bf E}^{*}_i)\delta_s({\bf r}). \label{eq_april25_2025_1}
\end{eqnarray}
The extinction power is then given again by eq.(\ref{eq_april25_2025_2}) with $w$ defined by eqs.(\ref{eq_april22_2025_2},\ref{eq_april22_2025_3}), but unlike in the prior section where the optical theorem expression is exact, in this section it is an approximation valid in the special case in which the source is far from the scatterer's region $\tau$. Then the sought-after expression for the extinction power can be written, e.g., as 
\begin{equation}
    P_e =\frac{1}{2} \Re \int_{\partial V} dS \left [ {\bf E}_i^*\cdot (\hat{\bf n}\times {\bf H}_s) - {\bf H}_i^*\cdot (\hat{\bf n}\times {\bf E}_s) \right ].
    \label{eq_april25_2025_10}
\end{equation}

Now, if the sensing domain $\partial V$ is in the far zone corresponding to the source then we can approximate these results further. In particular, if $\partial V$ is a large spherical volume of radius $R$ that is centered at the origin, then by choosing $\hat{\bf n}\rightarrow \hat{\bf r}$ and using the fact that for far fields ${\bf H}_i = \hat{\bf r}\times {\bf E}_i/\eta$ and ${\bf E}_i=-\eta\hat{\bf r}\times {\bf H}_i$ where $\eta$ denotes the free space impedance, then we obtain from eqs.(\ref{eq_april22_2025_2},\ref{eq_april22_2025_3})
\begin{equation}
    w=\frac{1}{\eta}\int_{\partial V} dS \left [ {\bf E}_s \cdot (\hat{\bf r}\times \hat{\bf r}\times {\bf E}_i^*) - (\hat{\bf r}\times {\bf E}_s) \cdot (\hat{\bf r}\times {\bf E}_i^*) \right ] .\label{eq_april25_2025_4}
\end{equation}
Moreover, if the surface $\partial V$ is also in the far zone of the scatterer's support $\tau$ then eq.(\ref{eq_april25_2025_4}) can be simplified further. In this case, both the incident fields and the scattered fields are perpendicular to $\hat{\bf r}$ in $\partial V$ and therefore in view of a well-known identity 
$\hat{\bf r}\times \hat{\bf r}\times {\bf E}_i^*= - (1 - \hat{\bf r}\hat{\bf r}\cdot) {\bf E}_i^*$ so that from (\ref{eq_april25_2025_4})
\begin{equation}
    w=-\frac{2}{\eta} \int_{\partial V} dS({\bf E}_i^* \cdot {\bf E}_s), \label{eq_april25_2025_5}
\end{equation}
and important result for practical implementations of the optical theorem in the far zone. Introducing the far-field radiation patterns ${\bf f}_i(\hat{\bf r})$ and ${\bf f}_s(\hat{\bf r})$ where 
\begin{eqnarray}
    {\bf E}_i({\bf r}) & \sim & {\bf f}_i(\hat{\bf r}) \frac{e^{ikr}}{r} \nonumber \\
    {\bf E}_s({\bf r}) & \sim & {\bf f}_s(\hat{\bf r}) \frac{e^{ikr}}{r} \label{eq_april25_2025_7}\end{eqnarray}
    as $kr\rightarrow\infty$
then one readily obtains from (\ref{eq_april25_2025_5}) the following expression: 
\begin{equation}
    w=-\frac{2}{\eta} \int_{4\pi} d\hat{\bf r} {\bf f}_i^*(\hat{\bf r})\cdot {\bf f}_s(\hat{\bf r}) \label{eq_april25_2025_8}\end{equation}
    where integration is over the unit sphere ($4\pi$ steradians). The corresponding extinction power is given from (\ref{eq_april25_2025_2}) by 
    \begin{equation}
        P_e = -\frac{1}{\eta} \Re \int_{4\pi} d\hat{\bf r} {\bf f}_i^*(\hat{\bf r})\cdot {\bf f}_s(\hat{\bf r}). \label{eq_april25_2025_11}
    \end{equation}

\subsection{Planar Aperture Realizations}

\begin{figure}
    \centering
    \includegraphics[width=0.9\linewidth]{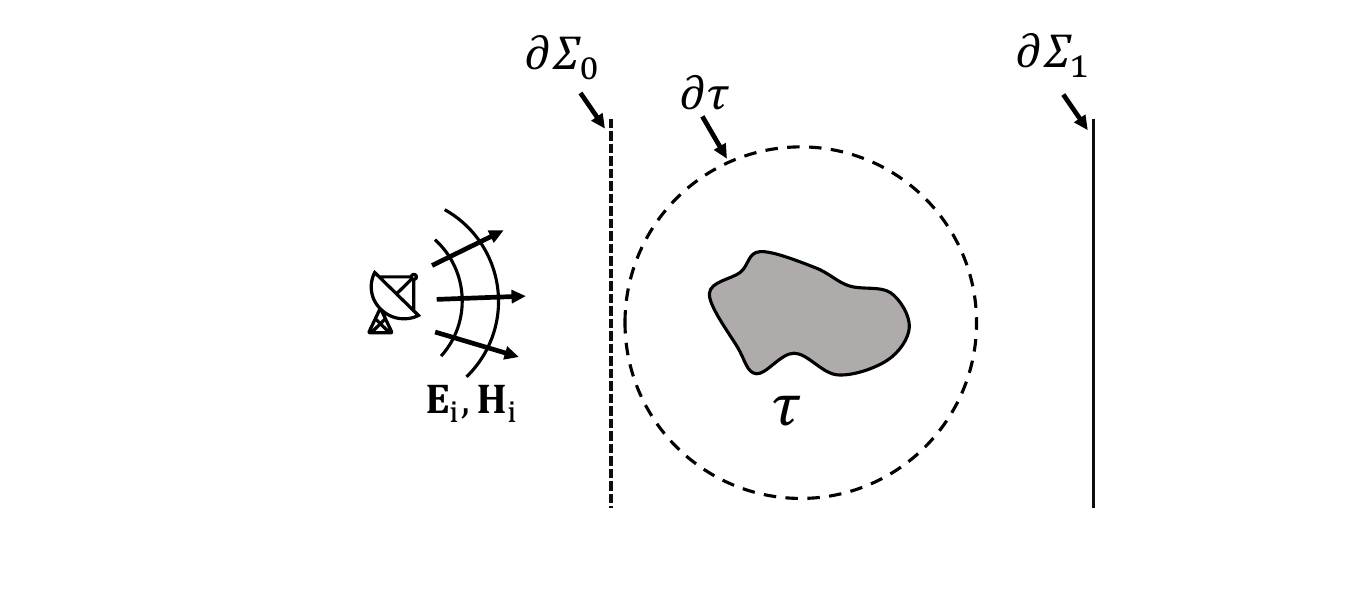}
    \caption{Planar aperture realizations involving either a single sensing plane ($\partial \Sigma_1$) or, more generally, two sensing planes ($\partial\Sigma_0$ and $\partial\Sigma_1$). }
    \label{fig_two_plane}
\end{figure}

An important practical realization of these developments is given by a planar sensor placed at a given sensing plane, say the surface $\partial \Sigma_1$ defined by $z=z_1$, in front of the scatterer, i.e., in the forward direction relative to the probing source, as shown in Figure~\ref{fig_two_plane}. It is assumed that the scattering region or ROI is inside the strip defined by $z_0\leq z\leq z_1$ and the source is in the half space defined by $z\leq z_0$. This sensor configuration is effective for the field-synthesis, in emission, of the c.c. probing field in this ROI, as long as the bounding plane $\partial \Sigma_0$ associated to $z=z_0$ is sufficiently far (typically a few wavelengths away) from the source. This ensures that the relevant probing field lacks significant evanescent content and its c.c. form is thereby reproducible, approximately, in this ROI, from $\partial \Sigma_1$, via conventional backpropagation. In this situation we get (in a form analogous to eqs.(\ref{eq_april25_2025_1}) and (\ref{eq_april25_2025_10}), after the substitution $\hat{\bf n}\rightarrow\hat{\bf z}$) that the relevant OT detector and corresponding expression for the extinction power are given by  
\begin{eqnarray}
 {\bf J}_s^{(OT)} &=& (\hat{\bf z}\times {\bf H}^{*}_i)\delta(z-z_1) \nonumber \\
    {\bf M}_s^{(OT)} &=& (\hat{\bf z}\times {\bf E}^{*}_i)\delta(z-z_1)
    \label{eq_june6_2025_1}
\end{eqnarray}
and
\begin{equation}
    P_e =\frac{1}{2} \Re \int_{\partial \Sigma_1} dS \left [ {\bf E}_i^*\cdot (\hat{\bf z}\times {\bf H}_s) - {\bf H}_i^*\cdot (\hat{\bf z}\times {\bf E}_s) \right ]
    \label{eq_april25_2025_10c1}
\end{equation}
or, equivalently (see eqs.(\ref{eq_april25_2025_2},\ref{eq_april22_2025_3})), 
\begin{equation}
    P_e =- \frac{1}{2}\Re\int_{\partial \Sigma_1} dS \left [ \hat{\bf z} \cdot \left ( {\bf E}_s \times {\bf H}_i^* + {\bf E}_i^* \times {\bf  H}_s \right ) \right ].
    \label{eq_june6_2025_8}
\end{equation}

Moreover, given the specialized planar form of the sensing surface, the following equivalent forms of OT detectors are also valid. In particular, it follows from image theory (\cite{balanis2024balanis}, page 331) that a pure ``electric-field-only'' planar optical theorem sensor defined by
\begin{eqnarray}
 {\bf J}_s^{(OT)} &=& 2(\hat{\bf z}\times {\bf H}^{*}_i)\delta(z-z_1) \nonumber \\ 
    {\bf M}_s^{(OT)} &=& 0
    \label{eq_june6_2025_2}
\end{eqnarray}
is equally valid, for which the corresponding extinction power expression is 
\begin{equation}
    P_e =-\Re \int_{\partial \Sigma_1} dS \left [ \hat{\bf z}\cdot \left ( {\bf E}_s\times {\bf H}_i^* \right)  \right ].
    \label{eq_april25_2025_10d1}
\end{equation}
Likewise, a pure ``magnetic-field only'' planar optical theorem sensor given by
\begin{eqnarray}
 {\bf J}_s^{(OT)} &=& 0 \nonumber \\
    {\bf M}_s^{(OT)} &=& 2(\hat{\bf z}\times {\bf E}^{*}_i)\delta(z-z_1)
    \label{eq_june6_2025_3}
\end{eqnarray}
is also valid, for which the corresponding extinction power is 
\begin{equation}
    P_e =-\Re \int_{\partial \Sigma_1} dS \left [ \hat{\bf z}\cdot \left (  {\bf E}_i^*\times {\bf H}_s \right ) \right ] .
    \label{eq_april25_2025_10e1}
\end{equation}
Importantly, in these developments, the scatterer can be close to the sensing surface; thus the scattered field can have arbitrary evanescent components in that plane. 

Another possibility is to employ two sensing planes, a forward plane ($\partial \Sigma_1$) and a backward one at $\partial\Sigma_0$, for the generation of the OT detector (see Fig.~\ref{fig_two_plane}). 
In this case the relevant c.c. probing field can be reconstructed exactly in the ROI, including evanescent components. Thus this OT detector applies even if the source is in the vicinity of bounding plane $\partial\Sigma_0$. The corresponding OT detector is such that when it acts as radiator it generates the c.c. probing field and is given (in a form analogous to eqs.(\ref{eq_june6_2025_1},\ref{eq_april25_2025_10c1},\ref{eq_june6_2025_8})) by 
\begin{eqnarray}
 {\bf J}_s^{(OT)} &=& (\hat{\bf z}\times {\bf H}^{*}_i)\delta(z-z_1)-(\hat{\bf z}\times {\bf H}^{*}_i)\delta(z-z_0) \nonumber \\
    {\bf M}_s^{(OT)} &=& (\hat{\bf z}\times {\bf E}^{*}_i)\delta(z-z_1)-(\hat{\bf z}\times {\bf E}^{*}_i)\delta(z-z_0)
    \label{eq_june6_2025_1b}
\end{eqnarray}
and
\begin{equation}
    P_e =-\frac{1}{2} \Re \sum_{m=0,1} (-1)^{m+1}\int_{\partial \Sigma_m} dS \left [ \hat{\bf z} \cdot \left ( {\bf E}_s\times {\bf H}_i^* + {\bf E}_i^*\times {\bf H}_s \right ) \right ].
    \label{eq_april25_2025_10c1b}
\end{equation}

\subsection{Multipole Domain Realization: Spherical Scanning}

We consider next the multipole domain realization of the OT detectors in both three-dimensional (3D) and two-dimensional (2D) spaces. These results are relevant for spherical scanning and cylindrical scanning geometries, respectively. Two forms of multipolar OT detectors are developed, which can be thought of as the multipole domain versions of the general results in subsections \ref{subsection_backpropagation} and \ref{subsection_surface_source}, respectively. We provide in this section the detailed derivation for the spherical geometry results. The corresponding results in the cylindrical coordinate system are elaborated in detail in the following section. 

\begin{figure}
    \centering
    \includegraphics[width=1\linewidth]{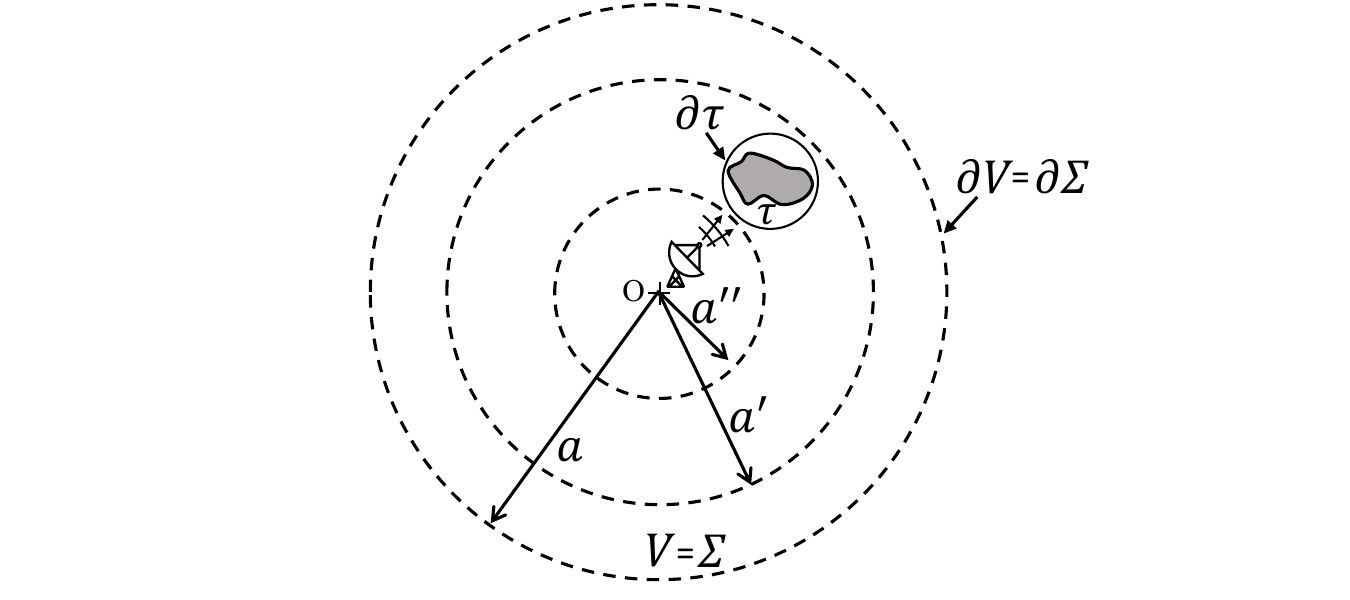}
    \caption{Topology for the backpropagation-based multipolar OT detector.}
    \label{fig_multipolar_case3}
\end{figure}

We consider first the  approach based on the adoption, as the sensing surface $\partial V$ in subsection \ref{subsection_backpropagation}, of a spherical surface of radius $a$, say $\partial \Sigma = \{ {\bf r}\in R^3| r= a\}$ which contains both the probing source and the scatterer, as shown in Fig.~\ref{fig_multipolar_case3}. The probing electric field outside a spherical volume centered about the origin that contains the source (say, of radius $a''$, as shown in the figure) can be expressed in the multipole expansion form: 
\begin{equation}
    {\bf E}_i({\bf r})=\sum_{l=1}^{\infty} \sum_{m=-l}^{l} A_{l,m} \nabla\times [h_l^{(+)}(kr){\bf Y}_{l,m}(\hat{\bf r})] + ik B_{l,m}h_l^{(+)}(kr){\bf Y}_{l,m}(\hat{\bf r}) %\label{eq_may8_2025_1}
\end{equation}
where $h_l^{(+)}(\cdot)$ is the spherical Hankel function of the first kind of order $l$, ${\bf Y}_{l,m}(\cdot)$ is the vector spherical harmonic of degree $l$ and order $m$, which is defined by 
\begin{equation}
    {\bf Y}_{l,m}(\hat{\bf r}) = {\bf L}Y_{l,m}(\hat{\bf r}) \label{eq_may9_2025_2}
\end{equation}
where ${\bf L}$ is the angular momentum operator and $Y_{l,m}(\cdot)$ is the ordinary spherical harmonic of degree $l$ and order $m$, and $A_{l,m}$ and $B_{l,m}$ are the electric and magnetic multipole moments, respectively, corresponding to the incident field probing the scatterer. 
Similarly, we can expand the associated scattered field, everywhere outside a spherical volume that is centered at the origin and contains the scatterer (of radius $a'>a''$, as shown in the figure), in the form: 
\begin{equation}
    {\bf E}_s({\bf r})=\sum_{l=1}^{\infty} \sum_{m=-l}^{l} a_{l,m} \nabla\times [h_l^{(+}(kr){\bf Y}_{l,m}(\hat{\bf r})] + ik b_{l,m}h_l^{(+)}(kr){\bf Y}_{l,m}(\hat{\bf r}) %\label{eq_may8_2025_1}
\end{equation}
where $a_{l,m}$ and $b_{l,m}$ represent the corresponding electric and magnetic multipole moments of the scattered field. 

It follows from the large argument approximation for the spherical Hankel function, and the properties of the vector spherical harmonics, that these fields behave in the far zone (as $ka\rightarrow \infty$) as indicated in eq.(\ref{eq_april25_2025_7}) with the replacement $r\rightarrow a$ where the far-field radiation patterns of the incident and scattered electric field are given by 
    \begin{eqnarray}
        {\bf f}_i(\hat{\bf r}) &=& \sum_{l,m} (-i)^l \left [ A_{l,m}\hat{\bf r}\times {\bf Y}_{l,m}(\hat{\bf r}) + B_{l,m} {\bf Y}_{l,m}(\hat{\bf r})\right ] \nonumber \\
        {\bf f}_s(\hat{\bf r}) &=& \sum_{l,m} (-i)^l \left [ a_{l,m}\hat{\bf r}\times {\bf Y}_{l,m}(\hat{\bf r}) + b_{l,m} {\bf Y}_{l,m}(\hat{\bf r})\right ] .
        \label{eq_may8_2025_3}\end{eqnarray}
    Substituting this result in (\ref{eq_april25_2025_11}), while using the orthogonality property of the vector spherical harmonics, one readily obtains 
    the sought-after expression for the extinction power: 
\begin{equation}
    P_{e} = -\frac{1}{\eta} \Re \sum_{l,m} (l)(l+1)\left ( a^m_l A^{m*}_l+ b^m_l B^{m*}_l \right ) .\label{eq_may8_20205_5}
\end{equation}

\begin{figure}
    \centering
    \includegraphics[width=0.9\linewidth]{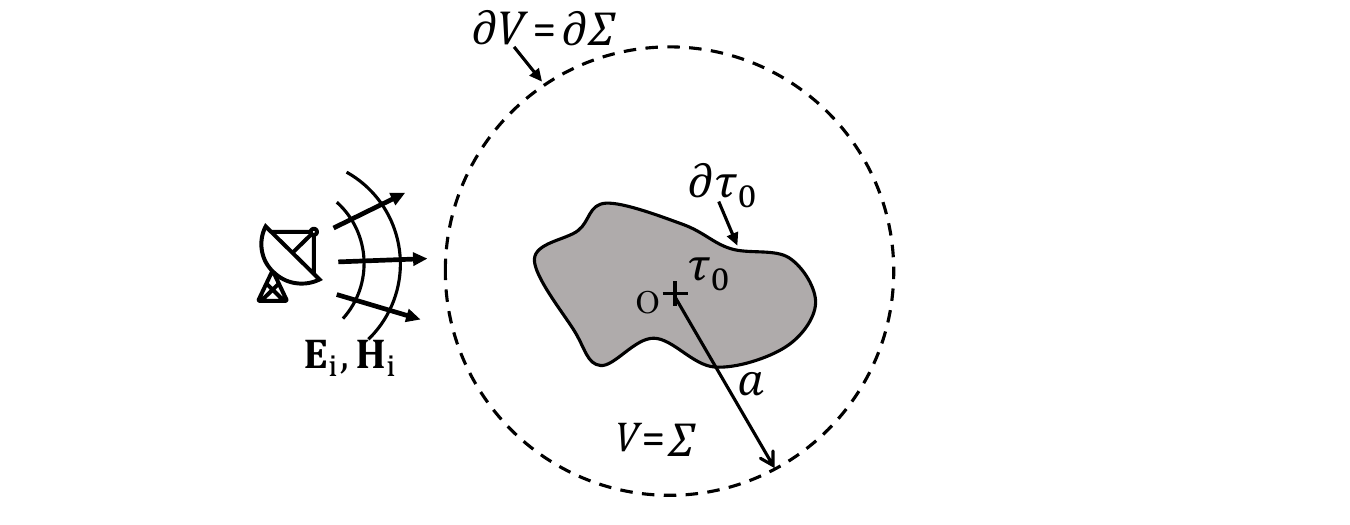}
    \caption{Topology for the general multipolar OT detector which applies in the near field region.}
    \label{fig_multipolar_case2}
\end{figure}

A second form of multipolar OT detector is obtained from the application of the results in subsection \ref{subsection_surface_source} to a spherical sensing surface. Figure~\ref{fig_multipolar_case2} illustrates the corresponding geometry. 
Let $\partial V = \partial \Sigma$. The most general incident electric field in the spherical volume $V=\Sigma = \{ {\bf r}\in R^3 | r\leq a\}$ bounded by this sensing domain is of the form 
\begin{equation}
{\bf E}_i({\bf r})=\sum_{l=1}^{\infty} \sum_{m=-l}^{l} C_{l,m} \nabla\times [j_l(kr){\bf Y}_{l,m}(\hat{\bf r})] + ik D_{l,m}j_l(kr){\bf Y}_{l,m}(\hat{\bf r}) \label{eq_may8_2025_7}
\end{equation}
where $j_l(\cdot)$ is the spherical Bessel function of the first kind of order $l$. The corresponding c.c. field for $r\leq a$ is then given by 
\begin{equation}
{\bf E}_i^*({\bf r})=\sum_{l=1}^{\infty} \sum_{m=-l}^{l} C_{l,m}^* \nabla\times [j_l(kr){\bf Y}_{l,m}^*(\hat{\bf r})] - ik D_{l,m}^*j_l(kr){\bf Y}_{l,m}^*(\hat{\bf r}) .\label{eq_may8_2025_7b}
\end{equation}
The corresponding magnetic field ${\bf H}_i$ for $r\leq a$ can be found from (\ref{eq_may8_2025_7}) and Maxwell's equations (\ref{eq_may24_2025_1}) to be given by
\begin{equation}
{\bf H}_i = \frac{1}{i\omega \mu} \sum_{l,m} k^2 C_{l,m} j_l(kr) {\bf Y}_{l,m}(\hat{\bf r}) + ik D_{l,m} \nabla \times [j_l(kr) {\bf Y}_{l,m}(\hat{\bf r})]
    \label{eq_may9_2025_1}
\end{equation}
where we have used the well-known identity $\nabla\times\nabla=\nabla\nabla\cdot -\nabla^2$ and $\nabla^2 [j_l(kr){\bf Y}_{l,m}(\hat{\bf r})]=-k^2 j_l(kr){\bf Y}_{l,m}(\hat{\bf r})$ as well as $\nabla\cdot [j_l(kr){\bf Y}_{l,m}(\hat{\bf r})]=0$. Now, it can be shown that (see, e.g., \cite{marengo2000new}, p.~862) 
\begin{equation}
    \nabla\times [j_l(kr){\bf Y}_{l,m}(\hat{\bf r})]=i\frac{l(l+1)}{r}j_l(kr)\hat{\bf r}Y_{l,m}(\hat{\bf r})+\frac{1}{r} \frac{d}{dr} \left [ r j_l(kr)\right ] \hat{\bf r}\times {\bf Y}_{l,m}(\hat{\bf r}) \label{eq_may9_2025_4}
\end{equation}
which enables us to rewrite eqs.(\ref{eq_may8_2025_7}) and (\ref{eq_may9_2025_1}) in the following forms, which facilitate application of the general result in eqs.(\ref{eq_april25_2025_2},\ref{eq_april22_2025_2},\ref{eq_april22_2025_3}):  
\begin{equation}
{\bf E}_i({\bf r})=\sum_{l,m} ikD_{l,m}j_l(kr){\bf Y}_{l,m}(\hat{\bf r}) + C_{l,m}\left  \{ i\frac{l(l+1)}{r} j_l(kr) \hat{\bf r}  Y_{l,m}(\hat{\bf r}) + \frac{1}{r} \frac{d}{dr} \left  [ r j_l(kr) \right ] \hat{\bf r}\times {\bf Y}_{l,m}(\hat{\bf r}) \right \} 
    \label{eq_may9_2025_12}
\end{equation}
and
\begin{equation}
{\bf H}_i({\bf r}) = \frac{1}{i\omega \mu} \sum_{l,m} k^2 C_{l,m} j_l(kr) {\bf Y}_{l,m}(\hat{\bf r}) + ik D_{l,m} \left  \{ i\frac{l(l+1)}{r} j_l(kr) \hat{\bf r}  Y_{l,m}(\hat{\bf r}) + \frac{1}{r} \frac{d}{dr} \left  [ r j_l(kr) \right ] \hat{\bf r}\times {\bf Y}_{l,m}(\hat{\bf r}) \right \} .\label{eq_may9_2025_8}
\end{equation}

Similarly, the scattered electric and magnetic fields admit the following representation for $r\geq a$:
\begin{eqnarray}
{\bf E}_s({\bf r}) &=& \sum_{l=1}^{\infty}\sum_{m=-l}^{l} F_{l,m} \nabla\times [h_l^{(+)}(kr){\bf Y}_{l,m}(\hat{\bf r})]+ikG_{l,m} h_l^{(+)}(kr) {\bf Y}_{l,m}(\hat{\bf r}) \nonumber \\
&=& \sum_{l,m} ikG_{l,m}h_l^{(+)}(kr) {\bf Y}_{l,m}(\hat{\bf r}) + F_{l,m} \left \{ i\frac{l(l+1)}{r}h_l^{(+)}(kr)\hat{\bf r}Y_{l,m}(\hat{\bf r}) + \right.\nonumber \\
&&\left. \frac{1}{r} \frac{d}{dr} \left[ r h_l^{(+)}(kr)\right ]  \hat{\bf r}\times {\bf Y}_{l,m}(\hat{\bf r}) \right \} \label{eq_may10_2025_1}\end{eqnarray}
and 
\begin{eqnarray}
{\bf H}_s({\bf r}) &=&\frac{1}{i\omega \mu} \sum_{l,m} k^2 F_{l,m}  h_l^{(+)}(kr) {\bf Y}_{l,m}(\hat{\bf r}) + ik G_{l,m} \left \{ i\frac{l(l+1)}{r} h_l^{(+)}(kr) \hat{\bf r} Y_{l,m}(\hat{\bf r}) + \right. \nonumber \\
&&\left. \frac{1}{r}\frac{d}{dr}\left [ r h_l^{(+)}(kr)\right ] \hat{\bf r}\times {\bf Y}_{l,m}(\hat{\bf r}) \right \}
    \label{eq_may10_2025_2}
\end{eqnarray}
where the expression for the magnetic field is given in a form analogous to the incident field counterpart eq.(\ref{eq_may9_2025_8}). 

Now, it follows from (\ref{eq_may9_2025_8}) and (\ref{eq_may10_2025_1}) and the orthogonality of the spherical harmonics $\hat{\bf r}Y_{l,m}(\hat{\bf r})$, ${\bf Y}_{l,m}(\hat{\bf r})$, and $\hat{\bf r}\times {\bf Y}_{l,m}(\hat{\bf r})$ that the first term in the integral in eq.(\ref{eq_april22_2025_3}), in particular, the quantity defined by 
\begin{equation}
    w_1=-\int_{\partial \Sigma} dS \hat{\bf n} \cdot \left [ {\bf E}_s({\bf r}) \times {\bf H}_i^{*}({\bf r}) \right ] = -a^2\int_{4\pi} d\hat{\bf r} \hat{\bf r}\cdot \left [ {\bf E}_s({\bf r}) \times {\bf H}_i^{*}({\bf r}) \right ]  
    \label{eq_may10_2025_4b}
\end{equation}
is given by
\begin{eqnarray}
w_1 &=&\sum_{l,m} \frac{l(l+1)(ka)^2}{i\omega \mu}\left \{ D_{l,m}^*G_{l,m} h_l^{(+)}(ka) \frac{1}{a} \frac{d}{dr} \left [ r j_l(kr) \right ]|_{r=a} \right. - \nonumber \\ 
&& \left. C_{l,m}^*F_{l,m} j_l(ka) \frac{1}{a} \frac{d}{dr} \left [r h_l^{(+)}(kr) \right ]|_{r=a} \right \}.
    \label{eq_may10_2025_5b}
\end{eqnarray}
Similarly, it follows from (\ref{eq_may9_2025_12}) and (\ref{eq_may10_2025_2}) that the second term in the integral in eq.(\ref{eq_april22_2025_3}), in particular, the quantity  defined by  
\begin{eqnarray}
  w_2= -\int_{\partial \Sigma} dS \hat{\bf n} \cdot \left [ {\bf E}_i^*({\bf r}) \times {\bf H}_s({\bf r}) \right ] &=& -a^2\int_{4\pi} d\hat{\bf r} \hat{\bf r}\cdot \left [ {\bf E}_i^*({\bf r}) \times {\bf H}_s({\bf r}) \right ]  
    \label{eq_may10_2025_4}
\end{eqnarray}
is given by 
\begin{eqnarray}
w_2 &=& -\sum_{l,m}\frac{l(l+1)(ka)^2}{i\omega \mu}\left \{ D_{l,m}^*G_{l,m} j_l(ka) \frac{1}{a} \frac{d}{dr} \left [ r h_l^{(+)}(kr) \right ]|_{r=a} \right. - \nonumber \\ 
&& \left. C_{l,m}^*F_{l,m} h_l(ka) \frac{1}{a} \frac{d}{dr} \left [r j_l(kr) \right ]|_{r=a} \right \}.
    \label{eq_may10_2025_5}
\end{eqnarray}
Then from (\ref{eq_may10_2025_5b}) and (\ref{eq_may10_2025_5}) we readily find, after some simplifications involving well-known expressions for the Wronskian for Bessel functions, that the quantity $w=w_1+w_2$ in eq.(\ref{eq_april22_2025_3}) can be expressed compactly in terms of the multipole moments of the incident and scattered fields as 
\begin{eqnarray}
    w=-\frac{1}{\eta} \sum_{l=1}^{\infty}\sum_{m=-l}^l l(l+1)\left ( C_{l,m}^*F_{l,m}+D_{l,m}^*G_{l,m}\right ) \label{eq_may10_2025_9}
\end{eqnarray}
and consequently from (\ref{eq_april25_2025_2}) the extinction power is given in terms of the multipole moments by 
\begin{equation}
P_e=-\frac{1}{2\eta} \Re \sum_{l=1}^{\infty} \sum_{m=-l}^l l(l+1) \left ( C_{l,m}^*F_{l,m}+D_{l,m}^*G_{l,m} \right ).
    \label{eq_may10_2025_11}
\end{equation}

It is instructive to point out that the key result eq.(\ref{eq_may10_2025_11}) can also be derived in a somewhat simpler form directly from the solution of an associated inverse problem. In particular, it can be shown from the multipole expansion (see, e.g., \cite{collin1990field}, ch. 2, eqs. (116), (139), (161), (165), (166)) and classical inversion methods (see, e.g., \cite{marengo1999inverse,marengo2004inverse},  for an overview of the relevant inverse source problem and further references)  that the c.c. incident field in eq.(\ref{eq_may8_2025_7b}) can be generated for $r\leq a$ using the equivalent ``minimum energy'' (ME) surface source 
\begin{equation}
{\bf J}_{ME}({\bf r}) = \delta (r-a) \sum_{l,m} \left [ \frac{i}{ka^2 \eta h_l^{(+)}(ka)}  \right ]D_{l,m}^*{\bf Y}_{l,m}^*(\hat{\bf r}) - \left \{ \frac{1}{r} \frac{d}{dr}\left [ r h_l^{(+)}(kr)\right ]|_{r=a} \right \}^{-1}\frac{1}{a^2\eta}C_{l,m}^*\hat{\bf r}\times {\bf Y}_{l,m}(\hat{\bf r}) \label{eq_may19_2025_1}
    \end{equation}
    which thereby functions itself as yet another OT detector. In particular, it is not hard to show with the help of eq.(\ref{eq_may9_2025_4}) and a similar expression involving the spherical Hankel function (see, e.g., Appendix A in \cite{marengo2000new}), along with the orthogonality of the vector spherical harmonics, that the projection of the scattered field in (\ref{eq_may10_2025_1}) onto the ME source in (\ref{eq_may19_2025_1}), acting as OT detector, is given by eq.(\ref{eq_may10_2025_9}), which therefore renders the same expression eq.(\ref{eq_may10_2025_11}) for the extinction power, as desired. 
We discuss next the corresponding results in cylindrical coordinates.
    %The detailed derivation is given in Appendix A. 

    \subsection{Multipole Domain Realization: Cylindrical Scanning}

We consider next the multipole representation results in 2D space, within the conventional cylindrical coordinate system, with radial coordinate $\rho$, azimuthal angle $\phi$, and axial coordinate $z$. We assume next that there is no $z$ dependence. Let (${\bf r}=(\rho,\phi)$). The most general z-oriented field in the ROI defined by $\rho\leq a$ can be synthesized using $z$-oriented electric and magnetic line sources, 
\begin{eqnarray}{ J}_e({\bf r})&=&I_e(\phi)\delta(\rho-a) \nonumber \\
{ J}_m({\bf r})&=&I_m(\phi)\delta(\rho-a), \label{eq_may21_2025_18}
\end{eqnarray} 
respectively. We discuss next the electric multipoles associated with such electric sources. The contributions associated with the companion magnetic sources are of the same general form, and are added in the final steps of the derivation.

The most general 2D incident electric field for $\rho\leq a$ 
is of the form 
\begin{equation}
    E_i(\rho,\phi)=\sum_{n=-\infty}^{\infty} A_n J_n(k\rho) e^{i n \phi} \label{eq_may21_2025_1}
\end{equation}
where the expansion coefficients $A_n$ are the corresponding electric multipole moments.
Then the corresponding c.c. version of the incident field is 
\begin{equation}
    E_i^*(\rho,\phi)=\sum_{n=-\infty}^{\infty} A_n^* J_n(k\rho) e^{-in\phi} . \label{eq_may21_2025_8}
\end{equation}

To implement the optical theorem, we seek a source $J_e$ as defined in eq.(\ref{eq_may21_2025_18}) whose generated field in the ROI ($\rho\leq a$) is the c.c. field in eq.(\ref{eq_may21_2025_8}). The map from $J_e$ to this field is of the general form (see, e.g., \cite{balanis2024balanis}, sec. 11.2) 
\begin{equation}
    E_i^*(\rho,\phi)=-\frac{k^2}{4\omega\epsilon} \int d\rho' \rho' \int_{0}^{2\pi} d\phi' H_0^{(+)}(k|{\bf r}-{\bf r'}|)I_e(\phi')\delta(\rho'-a) ,
\end{equation}
which in view of the addition theorem for Hankel functions,
\begin{equation}
    H_0^{(+)}(k|{\bf r}-{\bf r'}|)=\sum_{n=-\infty}^{\infty} J_n(k\rho_<)H_n^{(+)}(k\rho_>)e^{in(\phi-\phi')}
    \label{eq_may21_2025_2}
\end{equation}
where $\rho_<={\rm min}(\rho,\rho')$ and $\rho_>={\rm max}(\rho,\rho')$ (see, e.g., \cite{jackson2021classical}, p. 85 ), and the properties $J_{-n}(x)=(-1)^n J_n(x)$, $H_{-n}^{(+)}(x)=(-1)^n H_n^{(+)}(x)$ for integer $n$, can be expressed in the form given in eq.(\ref{eq_may21_2025_8}) where the coefficients $A_n^*$ are related to the source $I_e$ by 
\begin{equation}
    A_n^* = -\frac{ak^2}{4\omega \epsilon} H_n^{(+)}(ka) \int_{0}^{2\pi}d\phi e^{in\phi} I_e(\phi). \label{eq_may21_2025_6}
\end{equation}
Now, the most general source $I_e$ can be expanded as \begin{equation}
    I_e(\phi)=\sum_{m=-\infty}^{\infty} \alpha_m e^{-im\phi}. \label{eq_may21_2025_10}
\end{equation}
Substituting this in (\ref{eq_may21_2025_6}) and using the orthogonality property of the complex exponential functions $e^{in\phi}$ we readily find that the required source multipole moments $\alpha_n$ are 
\begin{equation}
    \alpha_n=-\frac{2\omega\epsilon}{a k^2 \pi}\left [ H_n^{(+)}(ka) \right]^{-1} A_n^* . \label{eq_may21_2025_12}
\end{equation}
The electric source $J_e$ with $I_e$ defined as in eqs.(\ref{eq_may21_2025_10},\ref{eq_may21_2025_12}) generates the c.c. field $E_i^*$ in eq.(\ref{eq_may21_2025_8}) and is thus a valid OT detector. 

If a scatterer appears in the ROI, then an associated scattered field is generated, which admits for $\rho\geq a$ the general multipole expansion 
\begin{equation}
    E_s(\rho,\phi)=\sum_{n=-\infty}^{\infty} C_n H_n^{(+)}(k\rho)e^{in\phi} . \label{eq_may21_2025_15}
\end{equation}
The corresponding projective measurement $w_e$ of this field onto the OT detector source $J_e$, in particular,
\begin{equation}
    w_e=\int d\rho \rho \int_{0}^{2\pi}d\phi  J_e(\rho,\phi) E_s (\rho,\phi), \label{eq_may21_2025_21}
\end{equation}is given from eqs.(\ref{eq_may21_2025_18},\ref{eq_may21_2025_10},\ref{eq_may21_2025_12},\ref{eq_may21_2025_15}) by 
\begin{equation}
    w_e=-\frac{4}{k\eta} \sum_{n=-\infty}^{\infty} A_n^* C_n \label{eq_may21_2025_23}
\end{equation}
where we have used $\eta=\sqrt{\mu/\epsilon}$. 

For the magnetic multipoles, we consider incident and scattered magnetic fields of the forms
\begin{eqnarray}
    H_i(\rho,\phi)&=&\frac{1}{\eta} \sum_{n=-\infty}^{\infty} B_n J_n(k\rho) e^{in\phi} \nonumber \\ 
    H_s(\rho,\phi)&=&\frac{1}{\eta} \sum_{n=-\infty}^{\infty} D_n J_n(k\rho)e^{in\phi} \label{eq_may21_2025_30}
\end{eqnarray}
and we obtain after manipulations analogous to the ones given above for the electric multipoles that the sought-after expression for the OT detector in terms of the multipole moments in 2D space, including both the electric multipole fields as well as the magnetic multipole fields in eq.(\ref{eq_may21_2025_30}) is given by 
\begin{equation}
    P_e = \frac{1}{2} \Re w\label{eq_may21_2025_51}
\end{equation}
where $w$ is the total projective measurement including both electric and magnetic multipolar contributions and is given by
\begin{equation}
    w= -\frac{4}{k\eta}\sum_{n=-\infty}^{\infty} A_n^* C_n + B_n^* D_n . \label{eq_may21_2025_40}
\end{equation}

\subsection{Discussion}
We conclude the section with a discussion about some of the general features noticed in all of the derived 
optical theorems along with their potential applications. We begin by noting, within the preceding multipole theory in 2D space, that
according to Poynting's theorem, applied to the above scattered fields, and the large argument approximation for the Hankel function, 
the scattered power $P_s$ is equal to 
\begin{equation}
    P_s = \frac{2}{k\eta}\sum_{n=-\infty}^{\infty} |C_n|^2 + |D_n|^2, \label{eq_may21_2025_50}
\end{equation}
an incoherent sum of the electric and magnetic multipole moments.
If the scatterer is lossless, then $P_e=P_s$, and therefore from (\ref{eq_may21_2025_51},\ref{eq_may21_2025_40},\ref{eq_may21_2025_50}) we also find that  
\begin{equation}
-\Re \sum_{n=-\infty}^{\infty} A_n^*C_n+B_n^*D_n = \sum_{n=-\infty}^{\infty} |C_n|^2+|D_n|^2 ,
\label{eq_may21_2025_52}
\end{equation}
an interesting general relation that applies to any lossless scatterer in 2D space. 
 More generally, for a scatterer made of passive media, $ P_e \geq P_s$ (due to possible energy dissipation in the interior of  the scatterer), and therefore
\begin{equation}
-\Re \sum_{n=-\infty}^{\infty} A_n^*C_n+B_n^*D_n  \geq\sum_{n=-\infty}^{\infty} |C_n|^2+|D_n|^2 .
\label{eq_may21_2025_52b}
\end{equation}

Similar considerations apply to the 3D multipole representation results in eq.(\ref{eq_may10_2025_11}). In particular, it is not hard to show from Poynting's theorem that the corresponding scattered power is 
\begin{equation}
    P_s=\frac{1}{2\eta} \sum_{l=1}^{\infty}\sum_{m=-l}^{l} (l)(l+1)\left ( |F_{l,m}|^2+|G_{l,m}|^2\right )
    \label{eq_june7_2025_1}
\end{equation}
and therefore in a manner analogous to eqs.(\ref{eq_may21_2025_52},\ref{eq_may21_2025_52b})
for lossless scatterers $P_s$ in eq.(\ref{eq_june7_2025_1}) is equal to $P_e$ in (\ref{eq_may10_2025_11}) so that 
\begin{equation}
-\Re \sum_{l=1}^{\infty} \sum_{m=-l}^l l(l+1) \left ( C_{l,m}^*F_{l,m}+D_{l,m}^*G_{l,m} \right ) = \sum_{l=1}^{\infty}\sum_{m=-l}^{l} (l)(l+1)\left ( |F_{l,m}|^2+|G_{l,m}|^2\right ) \label{eq_june7_2025_2}
\end{equation}
while, more generally, for passive scatterers (for which $P_e\geq P_s$) 
\begin{equation}
-\Re \sum_{l=1}^{\infty} \sum_{m=-l}^l l(l+1) \left ( C_{l,m}^*F_{l,m}+D_{l,m}^*G_{l,m} \right ) \geq \sum_{l=1}^{\infty}\sum_{m=-l}^{l} (l)(l+1)\left ( |F_{l,m}|^2+|G_{l,m}|^2\right ).\label{eq_june7_2025_2b}
\end{equation}

Expressions (\ref{eq_may21_2025_52}) and (\ref{eq_june7_2025_2}) for lossless scatterers and expressions (\ref{eq_may21_2025_52b}) and (\ref{eq_june7_2025_2b}) for passive scatterers can be written compactly in a general (non-representation-specific) Dirac bra-ket notation form: 
\begin{eqnarray}
    -\Re < \hat \psi_i | \psi_s> &=& || \psi_s ||^2 \quad {\rm lossless} \nonumber \\ 
    -\Re <\hat \psi_i | \psi_s> & \geq & ||\psi_s ||^2 \quad {\rm passive}
    \label{eq_june7_2025_5}
\end{eqnarray}
where $|\psi_s>$ represents (in Dirac bra-ket notation) the relevant scattered field vector or ket,  $|\hat \psi_i>$ is a companion vector or ket associated with the incident field interrogating the scatterer, $||\cdot||$ represents 2-norm, and $<\hat \psi_i|\psi_s>$ is the (Hermitian) projection of $|\psi_s>$ onto $|\hat \psi_i>$. 

In the ``second'' multipole theories leading to eqs.(\ref{eq_may10_2025_11}) and (\ref{eq_may21_2025_51},\ref{eq_may21_2025_40}) the companion vector $|\hat \psi_i>$ has a different physical meaning (in other words, it lies in a different space, physically) than the scattered field vector $|\psi_s>$, since it represents a converging or incoming wave into the ROI while $|\psi_s>$ represents a diverging or outgoing wave. This contrasts with the results pertinent to the ``first'' multipole theory in eq.(\ref{eq_may8_20205_5}), which in turn is based on the backpropagation-based imaging approach in eq.(\ref{eq_april25_2025_11}). In those developments the relevant incident field descriptor $|\hat \psi_i>$ is precisely the incident field and in that case both $|\hat \psi_i>$ and $|\psi_s>$ represent outgoing waves. We discuss next the counterpart of eq.(\ref{eq_june7_2025_5}) for those alternative frameworks. 

It follows from Poynting's theorem that the scattered power is given in terms of the scattered field radiation pattern ${\bf f}_s(\hat{\bf r})$ by 
\begin{equation}
    P_s = \frac{1}{2\eta}\int_{4\pi} d\hat{\bf r} |{\bf f}_s(\hat{\bf r})|^2 = \frac{1}{2\eta} ||{\bf f}_s||^2 \label{eq_june7_2025_11}
\end{equation}
from which, in view of eq.(\ref{eq_may8_2025_3}) and the orthogonality of the vector spherical harmonics, we also obtain 
\begin{equation}
    P_s=\frac{1}{2\eta} \sum_{l=1}^{\infty} \sum_{m=-l}^{l} (l)(l+1) \left ( |a_{l,m}|^2 + |b_{l,m}|^2 \right ). \label{eq_june7_2025_14}
\end{equation}
Comparing eqs.(\ref{eq_june7_2025_11},\ref{eq_june7_2025_14}) with eqs.(\ref{eq_april25_2025_11},\ref{eq_may8_20205_5}) we readily conclude that the 
counterpart of eq.(\ref{eq_june7_2025_5}) for the backpropagation-based imaging framework is 
\begin{eqnarray}
-\Re <\psi_i|\psi_s> &=& \frac{1}{2} ||\psi_s||^2 \quad {\rm lossless} \nonumber \\ 
-\Re <\psi_i|\psi_s> &\geq & \frac{1}{2} ||\psi_s||^2 \quad {\rm passive} 
    \label{eq_june7_2025_16}
\end{eqnarray}
where in this case the incident field descriptor is expressed as $|\psi_i>$ to highlight that it corresponds to the incident field as measured, e.g., in the same sensing aperture adopted for the sensing of the corresponding scattered field, or (mathematically) residing in the same functional space adopted for the description of the scattered field. The same general result eq.(\ref{eq_june7_2025_16}) applies, of course, to the (diffraction-theory-based) single-plane realization of the OT detectors associated with eqs.(\ref{eq_june6_2025_8},\ref{eq_april25_2025_10d1},\ref{eq_april25_2025_10e1}). 

To conclude this section we briefly consider some of the practical implications of eq.(\ref{eq_june7_2025_16}) for the detection of targets or medium changes, with  particular interest in the context of the radiation-pattern-based expression eq.(\ref{eq_april25_2025_11}) where ${\bf f}_i(\hat{\bf r})$ plays the role of $|\psi_i>$ while ${\bf f}_s(\hat{\bf r})$ corresponds to $|\psi_s>$.  We see from eq.(\ref{eq_june7_2025_11}) as well as from the related multipole expansion form in eq.(\ref{eq_june7_2025_14}) that within the full-view sensing context (i.e., for all directions over the unit sphere, or for all the multipole mode indices $l,m$) the 2-norm of $|\psi_s>$ is directly proportional to the scattered power. Thus, for simplicity, next we regard the 2-norm of the field essentially as the scattered power.  
Now, taking the 2-norm of the sensed total field (incident plus scattered field), i.e., $|\psi_i>+|\psi_s>$, we get 
\begin{eqnarray}
||\psi_i+\psi_s||^2 &=&||\psi||^2+2\Re<\psi_i|\psi_s>+||\psi_s||^2    \nonumber \\
&\leq &||\psi_i||^2 \label{eq_june7_2025_20}
\end{eqnarray}
    where the inequality stems from eq.(\ref{eq_june7_2025_16}). This result has the physical interpretation that for lossless scatterers the incident field is perturbed, through the secondary radiations originating at the scatterer, but the energy is simply re-radiated and thus not surprisingly the 2-norm of the total field is preserved. For passive scatterers, on the other hand, some of the energy is lost inside the scatterer and therefore the net 2-norm of the field is reduced in the presence of the scatterer, relative to the (background) scenario without the scatterer. These considerations hold for the full-view scenario, but these  findings also give insight about fundamental bounds pertinent to the practical situation in which both the probing and scattered fields are sensed only in a spatially-limited aperture, i.e., within a limited angular domain (say, solid angle $\Omega <4\pi$) or for a limited number of sensing modes in whatever representation is adopted. Within this ``limited-view'' case, we consider the situation in which the ROI where scatterers or targets appear is within the region where the required c.c. probing fields can be effectively launched from  the  given sensing aperture. Mathematically, this can be expressed (from the optical theorem) as 
    \begin{equation}
        -\Re<\tilde \psi_i|\tilde \psi_s>\simeq -\Re<\psi_i|\psi_s> \label{eq_june7_2025_42}
    \end{equation}
where $|\tilde\psi_i>$ and $|\tilde\psi_s>$ represent the corresponding limited-view fields. Moreover, the corresponding 2-norm of the limited-view scattered field is of the general form 
\begin{equation}
    ||\tilde\psi_s||^2 = \frac{\Omega}{4\pi} D ||\psi_s||^2 < ||\psi_s||^2 \label{eq_june7_2025_43} 
\end{equation}
where $D$ represents the gain or directivity of the scattered beam into the sensing domain. For isotropic re-radiation where $D=1$ the locally accessible 2-norm is a fraction ($\Omega/4\pi$) of the scattered field 2-norm. Thus locally only a fraction of the scattered power is measurable through the locally accessible field 2-norm. In contrast, the OT detector implicit in eq.(\ref{eq_june7_2025_42}) captures (approximately) the total extinction power (both the locally accessible component as well as the complementary, remote component). This concept offers potential applications in the detection and localization of scatterers or targets \cite{marengo2024optical, marengo20244change, marengo20133optical}. 
These ideas can be formalized by noting that, in view of eqs.(\ref{eq_june7_2025_42},\ref{eq_june7_2025_43}) and eq.(\ref{eq_june7_2025_16}), the following fundamental bounds hold for this practical limited-view sensing scenario:
\begin{equation}
     \frac{1}{2\eta}||\tilde\psi_s||^2 < \frac{1}{2\eta}||\psi_s||^2 \leq P_e \simeq -\frac{1}{\eta}\Re <\tilde \psi_i|\tilde \psi_s> . \label{eq_june9_2025_1}
\end{equation}

\section{Conclusion}

We have demonstrated for homogeneous backgrounds such as free space the multiple ways in which power extinction of scatterers can be sensed through suitably selected field measurements in the exterior of the scattering region. This has led to several equivalent expresssions for the optical theorem, corresponding to different configurations or representational domains, including surface source realizations, backpropagation-based sensors, planar detectors, as well as multipolar detectors in both 2D and 3D domains for spherical and cylindrical scanning geometries, respectively. Moreover, even within each category, a number of alternative forms of such detectors, referred to in the paper as ``OT detectors'', have emerged. 

The nonunique nature of the associated OT detectors originates from the existence of sources that are equivalent with respect to the scattering region, in the sense that they launch the same fields in said region. In particular, as shown in the paper, the synthesis of OT detectors relies on the solution of an associated inverse source problem, wherein the sought-after detectors must be such that they emit, in the complementary transmit or radiation mode, the c.c. form of the corresponding probing fields passing by the scattering region. The inverse source problem is well-known to be inherently nonunique, and this nonuniqueness is thereby passed to the sought-after detectors and their corresponding optical theorem expressions. Prior work \cite{marengo2015nonuniqueness} has demonstrated this equivalence of optical theorems for scalar fields. In the present paper we have expanded the research initiated in \cite{marengo2015nonuniqueness} by elaborating the corresponding equivalence for the full vector, electromagnetic framework.
We have explicitly highlighted, within the full vector theory, the myriad of forms in which optical-theorem-based detection can be implemented. 

We have  found that all of the derived OT detector versions have striking similarities. 
In particular, one of the most fundamental results that follows from these developments is the realization 
that all of the expressions for the extinction power consistently involve the
projection of a well-defined scattered field vector (e.g., in the form of the relevant far-field pattern of the scattered field, its multipole moments, or -~more directly~- the sensed fields as in the planar realizations) with a similarly well-defined reference vector. The latter is a vector that defines the probing field in the region under surveillance where scatterers can appear. In practical applications, this background field vector can be measured prior to the appearance of such targets. This facilitates the extrapolation of these ideas to other fields, such as target detection and estimation problems, facilitating the formulation of constraints associated with physical energy of the scattering phenomenon in an elegant, compact manner, in the form of a simple inner product (eq.(\ref{eq_june7_2025_16})) and companion bounds (eq.(\ref{eq_june9_2025_1})). The derived developments also facilitate understanding of the cross-field interactions when a target is probed by two or more superposition fields, with possible applications in intensity-only cross-field imaging. We are currently working in the latter direction and in other envisioned applications of the results derived in this paper including cloaking systems.

\bibliography{main}
\bibliographystyle{unsrt}

\end{document}